\input psfig.sty
\def\ptitle{Convexity and potential sums for Salpeter-like Hamiltonians} 
\def\ptitlea{Convexity and potential sums} 
\def\ptitleb{for Salpeter-like Hamiltonians} 
\nopagenumbers 
\magnification=\magstep1 
\hsize 6.0 true in 
\hoffset 0.25 true in 
\emergencystretch=0.6 in 
\vfuzz 0.4 in 
\hfuzz 0.4 in 
\vglue 0.1true in 
\mathsurround=2pt 
\def\nl{\noindent} 
\def\nll{\hfil\break\noindent} 
\def\np{\hfil\vfil\break} 
\def\ppl#1{{\noindent\leftskip 9 cm #1\vskip 0 pt}} 
\def\title#1{\bigskip\noindent\bf #1 ~ \trr\smallskip} 
 
\def\wbar#1{\overline{#1}} 
\font\trr=cmr10    
\font\bf=cmbx10    
\font\bmf=cmmib10  
\font\bmfs=cmmib8  
\font\sls=cmsl10 scaled 800 
\font\it=cmti10    
\font\trbig=cmbx10 scaled 1500 
\font\th=cmti10    
\font\tiny=cmr8    
\def\mb#1{\hbox{\bmf#1}}    
\def\mbs#1{\hbox{\bmfs#1}}  
\def\ng{>\kern -9pt|\kern 9pt} 
\def\hi#1#2{$#1$\kern -2pt-#2} 
\def\hy#1#2{#1-\kern -2pt$#2$} 

\def\sgn{{\rm sgn}} 
\def\half{{1 \over 2}} 
 
\def\nl{\noindent}    
\def\nll{\hfil\break} 
\def\dbox#1{\hbox{\vrule 
\vbox{\hrule \vskip #1\hbox{\hskip #1\vbox{\hsize=#1}\hskip #1}\vskip #1 
\hrule}\vrule}} 
 
\def\qed{\hfill \dbox{0.05true in}} 
\output={\shipout\vbox{\makeheadline\ifnum\the\pageno>1 {\hrule} \fi 
{\pagebody}\makefootline}\advancepageno} 
 
\headline{\noindent {\ifnum\the\pageno>1 
{\tiny \ptitle\hfil page~\the\pageno}\fi}} 
\footline{} 
\newcount\zz \zz=0 
\newcount\q 
\newcount\qq \qq=0 
 
\def\pref #1#2#3#4#5{\frenchspacing \global \advance \q by 1 
\edef#1{\the\q}{\ifnum \zz=1 { %
\item{[\the\q]}{#2} {\bf #3},{ #4.}{~#5}\medskip} \fi}} 
 
\def\bref #1#2#3#4#5{\frenchspacing \global \advance \q by 1 
\edef#1{\the\q}{\ifnum \zz=1 { %
\item{[\the\q]}{#2}, {\it #3} {(#4).}{~#5}\medskip} \fi}} 
 
\def\gref #1#2{\frenchspacing \global \advance \q by 1 
\edef#1{\the\q}{\ifnum \zz=1 { %
\item{[\the\q]}{#2}\medskip} \fi}}

\def\sref #1{~[#1]} 
 
\def\references#1{\zz=#1 
\parskip=2pt plus 1pt 
{\ifnum \zz=1 {\noindent \bf References \medskip} \fi} \q=\qq 

\pref{\bse}{E.~E.~Salpeter and H.~A.~Bethe, Phys.~Rev.}{84}{1232 (1951)}{} 
\pref{\se}{E.~E.~Salpeter, Phys.~Rev.}{87}{328 (1952)}{} 
\bref{\lieb}{E.~H.~Lieb and M.~Loss}{Analysis}{American Mathematical Society, 
New York, 1996} {The definition of the Salpeter kinetic-energy operator is 
given on p.~168.}
\pref{\hallsa}{R.~L.~Hall, W.~Lucha, and F.~F.~Sch\"oberl, J.~Math.~Phys.}{42}{5228 (2001)}{} 
\pref{\hallkp}{R. L. Hall, J. Math. Phys.}{25}{2708 (1984)}{}
\pref{\hallenv}{R. L. Hall, J. Math. Phys.}{34}{2779 (1993)}{} 
\pref{\hallsb}{R.~L.~Hall, W.~Lucha, and F.~F.~Sch\"oberl, Int.~J.~Mod.~Phys. A}
{17}{1931 (2002)}{}
\bref{\fel}{W. Feller}{An introduction to probability theory and its 
applications, Volume II}{John Wiley, New York, 1971} {Jensen's inequality is 
discussed on p.~153.} 
\pref{\hallprd}{R. L. Hall, Phys. Rev. D}{37}{540 (1988)}{}
\pref{\hallmix}{R. L. Hall, J. Math. Phys.}{33}{1710 (1992)}{}
\pref{\mart}{A. Martin and S. M. Roy, Phys. Lett. B}{233}{407 (1989)}{}
\gref{\herb}{I. W. Herbst, Commun. Math. Phys. {\bf 53}, 285 (1977);~{\bf 55},  316 (1977) (addendum).}
\pref{\weyl}{H. Weyl, Math. Ann.}{71}{441 (1911)}{}
\pref{\fan}{Ky Fan, Proc. Nat. Acad. Sci. (U.S.)}{35}{652 (1949)}{}
\bref{\wein}{A. Weinstein and W. Stenger}{Methods of Intermediate Problems for
 Eigenvalues}{Academic, New York, 1972}{Weyl's theorem is discussed on p. 163.}  
} 
 
\references{0} 
\topskip 2pt
\trr 
\ppl{CUQM-92}\ppl{HEPHY-PUB 755/02}\ppl{UWThPh-2002-12}\ppl{math-ph/0208042} 
\ppl{August 2002}\medskip 
\vskip 0.4 true in 
\centerline{\trbig \ptitlea}
\vskip 0.2 true in
\centerline{\trbig \ptitleb}
\vskip 0.4true in 
\baselineskip 12 true pt 
\centerline{\bf Richard L.~Hall$^1$, Wolfgang Lucha$^2$, and Franz 
F.~Sch\"oberl$^3$}\medskip 
\nll {\sls $^{1}$Department of Mathematics and Statistics, Concordia 
University, 1455 de Maisonneuve Boulevard West, Montr\'eal, Qu\'ebec, Canada 
H3G 1M8} 
\nll {\sls $^{2}$Institut f\"ur Hochenergiephysik, \"Osterreichische 
Akademie der Wissenschaften, Nikolsdorfergasse 18, A-1050 Wien, Austria} 
\nll {\sls $^{3}$Institut f\"ur Theoretische Physik, Universit\"at Wien, 
Boltzmanngasse 5, A-1090 Wien, Austria} 
  
\nll{\sls $^{1}$rhall@mathstat.concordia.ca, $^{2}$wolfgang.lucha@oeaw.ac.at, 
$^{3}$franz.schoeberl@univie.ac.at} 
\bigskip\medskip 
\baselineskip = 18true pt 
 
\centerline{\bf Abstract}\medskip 
\nl The semirelativistic Hamiltonian $H = \beta\sqrt{m^2 + p^2} + V(r),$ where $V(r)$ is a central
potential in $\Re^3,$ is concave in $p^2$ and convex in $p \equiv \sqrt{p^2}.$  This fact enables
us to obtain complementary energy bounds for the discrete spectrum of $H.$ By extending the
notion of `kinetic potential' we are able to find general energy bounds
on the ground-state energy $E$ corresponding to potentials with the form $V = \sum_{i}a_{i}f^{(i)}(r).$ In the case of
sums of powers and the $\log$ potential, where $V(r) = \sum_{q\ne 0} a(q)\sgn(q)r^q + a(0)\ln(r),$
the bounds can all be expressed in the semi-classical form
$$E\approx \min_{r}\left\{\beta\sqrt{m^2 + {1\over{r^2}}} + \sum_{q\ne 0} a(q)\sgn(q)(rP(q))^{q} + a(0)\ln(rP(0))\right\}.$$
\nl`Upper' and `lower' \hi{P}{numbers} are provided for $q = -1,1,2,$ and for the
$\log$ potential $q = 0.$  Some specific examples are discussed, to show the quality of the bounds.
\np 
\topskip 20pt
\title{1.~~Introduction} 
The Hamiltonian $H = K + V$ for the problems we study has the feature that either
the kinetic energy $K$ or the potential energy $V$ is nonlocal.  The most important example
is the semirelativistic ``spinless-Salpeter'' Hamiltonian\sref{\bse--\hallsa} given by
$$H = K(p) + V(r) = \sqrt{m^2 + p^2} + V(r),\eqno{(1.1)}$$
\nl where $r \equiv \|\mb{r}\|,\quad \mb{r}\in\Re^3,$ and $p \equiv \|\mb{p}\| \equiv \sqrt{\mb{p}^2}.$
  In this form at least $K(p)$ is nonlocal in configuration space and is defined as a multiplicative operator
in momentum space.  That is to say, $K\psi$ is defined to be what we get when
$\psi$ is transformed to momentum space, the multiplicative operator $K$ is applied,
and the result is transformed back to coordinate space.  Nonlocality is the main source of difficulty for this class of problems.  We study the discrete spectra of these Hamiltonians by
the use of approaches that make use of convexity and of spectral information already
obtained concerning related problems. In an earlier paper\sref{\hallsa} we studied the relationship between $H_0 = K + h$ and $H = K + g\circ h,$ where $g(h(r))$ is a smooth transformation
of a `base' potential $h(r).$  For cases in which $g$ had definite convexity, one could
then employ the so-called `kinetic-potential'\sref{\hallkp} formalism and `envelope theory'\sref{\hallenv} to 
construct upper or lower bounds to the discrete eigenvalues of $H$ by using the known
 spectrum of $H_0.$ 

  The present paper has two distinct aspects: we turn our attention firstly to the convexity of $K,$ as a function of $p$ or $p^2$; and then we look at potentials that are a {\it sum} of terms $V = \sum_{i}V^{(i)}.$  In Section~2 we extend the kinetic-potential formalism to include more general
kinetic-energy operators than the Schr\"odinger form $K = p^2$ studied earlier\sref{\hallkp}.  We have already found\sref{\hallsb} some implications of the fact that
$K$ is {\it concave} in $p^2.$  The spectral implications of the {\it convexity} of $K$ in $p$
 demand a new analysis. By using Jensen's inequality\sref{\fel} we are able to construct a framework in Section~3 which accommodates both cases.  Although it may not be immediately 
apparent, the treatment of potential {\it sums} also leads to an interesting convexity analysis of a completely different type\sref{\hallprd, \hallmix}.  This in turn yields an optimized general lower bound for the bottom of the spectrum expressed in terms of
 the kinetic potentials generated by `component' problems $K + V^{(i)}.$ We discuss this in 
general terms in Section~4. In the special case in which the component problems are pure-power potentials or the $\log$ potential, that is to say
$$V(r) =  \sum_{q\ne 0} a(q)\sgn(q)r^q + a(0)\ln(r),\eqno{(1.2)}$$
\nl the lower bound (Section~5)
and also upper bounds obtained by variational methods (Section~6) can all be expressed in terms of 
a semi-classical expression of the form
$$E\approx \min_{r}\left\{\sqrt{m^2 + {1\over{r^2}}} + \sum_{q\ne 0}a(q)\sgn(q)(rP(q))^{q} + a(0)\ln(rP(0))\right\}.\eqno{(1.3)}$$
\nl It is the goal of this paper to develop a general theory which leads to such a result and,
in particular, to determine \hi{P}{numbers} which guarantee that the approximation (1.3) is an upper or lower bound.  We shall find the appropriate upper and lower $P(q)$ for the cases $q = -1, 1, 2,$ and for the $\log$ potential $q = 0.$  In Section~7 we apply our general results to some specific examples.

Although we obtain very concrete results in the end, our study begins with a somewhat abstract
viewpoint.  We now make a few general remarks that will help motivate these starting considerations. An idea that runs through the work is one well known to those who study non-linear problems: we try to use transformations to make the most of any soluble problem that is at hand, or, at least,
one for which we have a good approximation.  The setting for our ideas is geometrical.  We suppose that we have an exact solution (or good bounds) for a `base' problem with Hamiltonian $\alpha p^2 + \beta h(r),$ and
we are interested in a Hamiltonian of the form $H = k(p^2) + g(h(r)),$ where $k$ and $g$ are monotone increasing smooth transformations.  It follows that the `tangent spaces' to $H$ 
are Schr\"odinger operators with the general form $H^{(\mbs{t})} = a + bp^2 + ch(r),$
where the parameters $\{a, b, c\}$ depend on the contact vector $\mb{t}.$  What we look for is
a theory that would allow us to deduce spectral information about $H$ from the `known' spectrum
of its tangents $H^{(\mbs{t})}.$ For example, if the transformation functions $k$ and $g$ are both concave, we would expect to obtain upper bounds via the  
spectral inequality  $H < H^{(\mbs{t})}.$  {\it Mutatis mutandis,} a complimentary theory is possible with operator tangent spaces of the form $H^{(\mbs{t})} = a + bp + ch(r),$ where
$K(p)$ is {\it convex} in $p$ and $g(h)$ is convex in $h:$ this leads to energy {\it lower} bounds.
Since we have already explored\sref{\hallsa} the
potential transformation $g,$ the main thrust of the present paper concerns kinetic-energy transformations with `base problems' respectively $p^2 + h$ and $p+h$ and to applications of the results when the potential in $H$ is a sum of terms. The more general case in which neither transformation $k(p)$ nor $g(h)$ is the identity follows immediately by combining the present and earlier results; hence we do not need to discuss this natural generalization in detail here.  Coulomb `components'  present a special difficulty for a lower bound with kinetic energy $p$ because the operator $p - v/r$ has no discrete eigenvalues.  However, for our main concern, the Salpeter Hamiltonian, we are able to make use of the very good Coulomb lower bound of Martin and Roy\sref{\mart} and so incorporate the Coulomb contribution smoothly into our general formulation.  

\title{2.~~Variety of kinetic potentials} 
The discrete eigenvalues of the self-adjoint operators we study may be characterized variationally.  Thus the bottom of the spectrum $E$ of $H = K + V$ is given by $E = \inf(\psi,H\psi),$ where the infimum is taken over all normalized functions in the domain ${\cal D}(H)\subset L^2(\Re^3).$   The idea behind kinetic potentials is to perform the minimization in two stages: we first find the constrained minimum $\wbar{V}(K; s)$ of $(\psi, V\psi),$ keeping the mean kinetic energy $(\psi, K\psi) = s$ constant; then, we recover $E$ by minimizing over the kinetic energy
 $s > 0.$ Thus we have
$$\wbar{V}(K; s) = \inf_{{{\scriptstyle \psi \in {\cal D}(H)} \atop {\scriptstyle (\psi,\psi) = 1}} \atop {\scriptstyle (\psi, K\psi) = s}} (\psi, V\psi)\quad \Rightarrow\quad E = \min_{s>0}\left\{s + \wbar{V}(K; s)\right\}.\eqno{(2.1)}$$
\nl We call the function $\wbar{V}(K;s)$ the kinetic potential of $V$ associated with the 
kinetic-energy operator $K;$ we shall write simply $\wbar{V}(s),$ if the kinetic-energy operator $K$ is fixed or is clear from the context. It follows immediately from the definition that the kinetic potentials absorb a positive coupling parameter in the sense that $\wbar{cV}(s)= c\wbar{V}(s).$ We note also that the elementary
\nll{\bf Comparison Theorem} 
$$\wbar{V}^{(1)}(s) < \wbar{V}^{(2)}(s)\quad \Rightarrow\quad E^{(1)} < E^{(2)}$$
follows immediately from (2.1).  The arguments we use are not restricted to dimension
 $N = 3:$ this choice allows us to illustrate the general results with some explicit well-known examples, without the distraction of the operator dependencies on $N.$

The reason for using this description of the spectral problem is that it lends itself to some interesting approximations.  Firstly, we have shown in the Schr\"odinger\sref{\hallkp} and Salpeter\sref{\hallsa} cases that, if $V(r) = g(h(r))$ and $g$ is monotone increasing and has definite convexity, then the approximation
$\wbar{V}(s) \approx g(\wbar{h}(s))$ leads to lower bounds if $g$ is convex and upper bounds if $g$ is concave.  In Section~4 of the present paper we shall extend to general $K$ the result obtained earlier\sref{\hallmix} for the Schr\"odinger case that kinetic potentials are subadditive, that is to say 
$$V(r) = h^{(1)}(r) + h^{(2)}(r)\quad\Rightarrow\quad\wbar{V}(s) \geq \wbar{h}^{(1)}(s) + \wbar{h}^{(2)}(s).\eqno{(2.2)}$$
\nl The lower energy bound then immediately follows from the above-mentioned comparison theorem for kinetic potentials. There is more to this result than meets the eye: it generates the optimum of a family of lower bounds; the details will be given in Section~4 below.  Extensions to sums with more than two terms (or, further, to mixtures generated by an integral) are immediate.  The principal limitation is that
each potential term alone, when added to the kinetic energy, must, for large enough coupling, support a discrete eigenvalue.
Thus $V(r) = -1/r + r$ is allowed but $V(r) = 1/r + r$ is not.

The `component' kinetic potential $\wbar{h}(s) = \wbar{h}(K; s)$ can be
constructed by use of a Legendre transformation from the eigenvalue function $E = F(v),$ in which $F(v)$ is the bottom of the spectrum of $H = K + vh(r),$ as a function of the coupling $v.$ In the Schr\"odinger case $H = p^2 + vh(r)$ we have shown that $F(v)$ is concave\sref{\hallkp} and moreover the kinetic potential for $K = p^2$ is given in terms of $F(v)$ by the transformation
$$s = F(v) - vF'(v),\quad \wbar{h}(K; s) = F'(v).\eqno{(2.3)}$$
\nl The concavity of the eigenvalue function $F(v)$has been proved for the Schr\"odinger case\sref{\hallkp} and the Salpeter case\sref{\hallsa} by the application of a simple variational argument.  By exactly similar reasoning we can show that the eigenvalue function for the operator $p + vh(r)$ is also concave in $v.$  Moreover, the Legendre transformation (2.3) is generic: it is valid for all kinetic-energy operators $K.$ This is an  immediate consequence of the concavity of $F(v),$ as the following equations clearly demonstrate:
$$F(v) = \min_{u > 0}\{F(u) - uF'(u) + v F'(u)\} =
 \min_{s > 0}\{s + v\wbar{h}(K; s)\}.$$
\nl Our principal assumption concerning $K$ is that it is at once a convex function of $p$ and a concave function of $p^2.$  This convexity is clearly true for our most important example, the relativistic kinetic energy $K = \sqrt{m^2 + p^2};$ however, we shall use this specific form only when we need to. 

We now turn from the general to some very specific results.  We shall need to have at our disposal some `component' kinetic potentials for the operators $K + v~\sgn(q)r^q,$ where $K = p$, or the Schr\"odinger case $K  = p^2.$ By elementary scaling arguments we can show that the dependence of the energy functions on the coupling $v$ are given by

$$ p + v\ \sgn(q)r^q\quad\Rightarrow\quad E = F^{(1)}(q;v) = F^{(1)}(q;1)v^{1\over{1+q}}\eqno{(2.4a)}$$
\nl and
$$p^2 + v\ \sgn(q)r^q\quad\Rightarrow\quad E = F^{(2)}(q;v) = F^{(2)}(q;1)v^{2\over{2+q}}.\eqno{(2.4b)}$$

\nl The Legendre transformation $F\leftrightarrow\wbar{h}$ given above in (2.3) now allows us to deduce the precise forms of the corresponding kinetic potentials.  For convenience we choose to write the kinetic potentials so obtained in a special way.  We change variables for the mean kinetic energy $s$ in the two cases respectively to $s = 1/r$ and $s = 1/r^2.$  It then follows from (2.3) by straightforward algebraic computations that the kinetic potentials for $h(r) = \sgn(q)r^q$ have similar convenient forms, namely
$$\wbar{h}(p;1/r) = \sgn(q)(P^{(1)}(q)r)^q\eqno{(2.5a)}$$
\nl and 
$$\wbar{h}(p^2;1/r^2) = \sgn(q)(P^{(2)}(q)r)^q,\eqno{(2.5b)}$$
\nl where the \hi{P}{numbers} are {\it defined} in terms of the $v= 1$ eigenvalues $E^{(i)}(q)= F^{(i)}(q;1),$ $i = 1,2,$ respectively 
by the explicit formulas
$$K = p\quad\Rightarrow\quad P^{(1)}(q) := \left|{{E^{(1)}(q)}\over{1+q}}\right|^{1+{1\over q}}|q|\ ,\quad q > -1,\  q\ne 0 \eqno{(2.6a)}$$
and
$$K = p^2\quad\Rightarrow\quad P^{(2)}(q) := \left|{{E^{(2)}(q)}\over{1+q/2}}\right|^{{1\over 2}+{1\over q}}\left|{q\over 2}\right|^{\half},\quad q > -2,\  q\ne 0.\eqno{(2.6b)}$$
\nl The energies are related to the kinetic potentials by specific realizations of the general formula (2.1): for example, we have in this spectral representation
$$p + v\ \sgn(q)r^q\quad\Rightarrow\quad E = \min_{r>0}\left\{{1\over r} + v\ \sgn(q)\left(P^{(1)}(q)r\right)^q\right\},\quad \ q > -1,\ q\neq 0.\eqno{(2.7)}$$
\nl One of our side goals is purely esthetic, namely we wish to end up with  `attractive formulas': after the changes of variable from $s$ to $r,$ the kinetic potentials look like the original power potentials themselves, but with the \hi{P}{factors} inserted.
\nl We turn now to the base potential $h(r) = \ln(r)$ and find by scaling arguments that
$$p + v \ln(r)\quad\Rightarrow\quad F^{(1)}(v) = vF^{(1)}(1)- v\ln(v)\eqno{(2.8a)}$$
\nl and
$$ p^2 + v \ln(r)\quad\Rightarrow\quad F^{(2)}(v) = vF^{(2)}(1)- \half v\ln(v).\eqno{(2.8b)}$$
\nl Consequently we obtain from the transformation (2.3) 
$$\wbar{h}(p;1/r) = \ln(P^{(1)}(0)r),\quad\wbar{h}(p^2;1/r^2) = \ln(P^{(2)}(0)r),\eqno{(2.9)}$$
\nl where 
$$P^{(1)}(0) = \exp(E^{(1)}(0)-1),\quad P^{(2)}(0) = {1\over{\sqrt{2}}}\exp\left(E^{(2)}(0)-\half\right). \eqno{(2.10)}$$

For the discussion of examples we shall need to have some specific $P$ values.  For the cases $q = -1,0,1,2$ we supply some of these numerical values in Table~1. This table has an eigenvalue symmetry because of the operator equivalence $p + r^2 \sim p^2 + r;$ it also has two omissions corresponding to $q = -1,$ because $p -1/r$ has no discrete eigenvalues.
We offer now a solution to this Coulomb difficulty. As we shall make clear in Section~4, viable Coulomb \hi{P}{numbers} are needed for {\it lower} bounds.  For our most important application $K = \sqrt{p^2 + m^2},$ a lower bound to the bottom of the spectrum 
of $H = K - v/r$ is provided by the Martin--Roy bound\sref{\mart}
$$E \ge e_L(v) = m\left({{1 + \sqrt{1-4v^2}}\over 2}\right)^{\half},\quad v < \half.\eqno{(2.11)}$$
\nl The condition $v < \half$ is a little more restrictive than the fundamental operator restriction $v < 2/\pi$: it was proved by Herbst\sref{\herb} that a Friedrichs extension exists for $H$ only if the Coulomb coupling is sufficiently small. The Coulomb lower bound has the same scaling law with respect to $m$ as does the exact energy: although $m$ originates in
the Hamiltonian inside the square root of the kinetic-energy term, it appears in the eigenvalue and in its lower approximation simply as an overall factor\sref{\hallsa}.  Now we construct a \hi{v}{dependent} \hi{P}{representation} for this lower bound.  We write (as a definition of 
$P_L(v)$)
$$e_L(v) = \min_{r}\left\{\sqrt{m^2 + {1\over{r^2}}} - {v\over{P_L(v)r}}\right\}.\eqno{(2.12)}$$
\nl An elementary calculation then shows that (2.11) and (2.12) imply $P_L(v) = e_L(v)/m.$  This serendipitous discovery fills the gaps in Table~1, and will allow us to include the Coulomb component in our lower-bound energy formula for sums of potential terms: we must make the substitution
$$-{v\over{P^{(1)}(-1)r}} = -{v\over{P_L(v)r}} = -{{mv}\over{e_L(v)r}},\quad v < \half.\eqno{(2.13)}$$     
 
\title{3.~~Complementary convexity: $p + V$ and $p^2 + V$} 
The principal result of this section is best expressed in terms of kinetic potentials by
the following
\nll{\bf Theorem~1.}~~
{\th If $E$ is the bottom of the spectrum of the Hamiltonian $H = K + V,$ and the kinetic-energy
 operator $K$ is at once convex in $p$ and
concave in $p^2,$  then it follows that}
$$\min_{s > 0}\left\{K(s) + \wbar{V}(p; s)\right\}\ \leq\  E\  \leq\  \min_{s > 0}\left\{K(s) + \wbar{V}(p^2; s^2)\right\}.\eqno{(3.1)}$$
\nl It makes sense here to speak of $K(p)$ as though $p$ were a real variable since,
by definition, the action of the operator $K$ is effected via the Fourier transform. 
We shall now prove this result by an application of Jensen's inequality\sref{\fel} and kinetic potentials defined in (2.1). We consider first the left-hand inequality of the theorem. If $\psi$ is a normalized function in the domain ${\cal D}(H)$ of $H = K + V,$
then, since $K = K(p)$ is convex in $p,$ by Jensen's inequality, we have
$$E = \inf_{{{\scriptstyle \psi \in {\cal D}(H)} \atop {\scriptstyle (\psi,\psi) = 1}}}
\left\{(\psi, K(p)\psi) + (\psi,V\psi)\right\} \geq \inf_{{{\scriptstyle \psi \in {\cal D}(H)} \atop {\scriptstyle (\psi,\psi) = 1}}}
\left\{K((\psi,p\psi)) + (\psi,V\psi)\right\}.$$
\nl That is to say,
$$ E \geq \min_{s >0}\inf_{{{\scriptstyle \psi \in {\cal D}(H)} \atop {\scriptstyle (\psi,\psi) = 1}} \atop {\scriptstyle (\psi, p\psi) = s}} \left\{K((\psi,p\psi)) + (\psi,V\psi)\right\}
= \min_{s >0}\left\{K(s) + \wbar{V}(p; s)\right\}.$$
\nl The proof of the upper-bound inequality is very similar: we write $K(p) = k(p^2),$ where
$k$ is concave; then, setting $t = s^2,$ we arrive at the inequality
$$E \leq \min_{t > 0}\left\{k(t) + \wbar{V}(p^2; t)\right\} = \min_{s > 0}\left\{K(s) + \wbar{V}(p^2; s^2)\right\},$$
\nl which establishes the theorem.\qed
\nll This result is an essential ingredient in the proof of the sum approximation in the next section.  We now look at an example, namely the Salpeter problem with a linear potential.
We have
$$H = \sqrt{m^2 + p^2} + V(r),\quad V(r) = v h(r) = v r,\eqno{(3.2)}$$
\nl where $v$ is a positive coupling parameter. In terms of the convenient variable $r > 0$ 
the two kinetic potentials from (2.5) are 
$$\wbar{h}(p;1/r) = P^{(1)}(1)r,\quad \wbar{h}(p^2;1/r^2) = P^{(2)}(1)r,\eqno{(3.3)}$$
\nl where the \hi{P}{numbers} are provided in Table~1.  Theorem~1 then immediately yields the
bounds
$$\min_{r >0}\left\{\sqrt{m^2 + r^{-2}} + v P^{(1)}(1)r\right\} \leq E \leq \min_{r >0}\left\{\sqrt{m^2 + r^{-2}} + v P^{(2)}(1)r\right\}.\eqno{(3.4)}$$
\nl In Figure~1 we plot these bounds as a function of $m$ for the case $v = 1$.
If we combine Theorem~1 here with Theorem~2 of Ref.\sref{\hallsa} (to the effect that
$\wbar{g\circ h} > g\circ\wbar{h}$ when $g$ is convex) we obtain the following
class of examples.  We suppose that $V(r)$ is monotone increasing and convex
in $h(r) = r$ then the two theorems together yield the lower bound
$$E \geq \min_{r >0}\left\{\sqrt{m^2 + {1\over{r^2}}} + v V(P^{(1)}(1)r)\right\}.\eqno{(3.5)}$$
\nl Of course, if $V(r)$ is {\it concave,} then we get an {\it upper} bound by the same expression
provided we use $P^{(2)}(1).$  It is perhaps important to note that 
with $P = P^{(2)}(r)$ an upper bound
would be obtained for every choice of $r$ in the expression on the
right-hand side; the 
expression in (3.5) is however only a lower bound {\it a priori} at the minimum point.
\title{4.~~The sum approximation: lower bounds} 
Since further generalization easily follows, we first look at the problem of the
sum of two potential terms. We assume that each potential $vh^{(i)}(r)$ alone, when added to the kinetic-energy operator $K,$ has a discrete eigenvalue at the bottom of the spectrum for sufficiently large `coupling' $v.$  We express our result in terms of kinetic potentials and prove the following
\nll{\bf Theorem~2}~~
{\th If $E$ is the bottom of the spectrum of the Hamiltonian $H = K + V,$ 
and the potential $V$ is the sum $V(r) = h^{(1)}(r) + h^{(2)}(r),$ then it follows that the sum of the component kinetic potentials yields a lower bound to $\wbar{V},$ that is to say}
$$\wbar{V}(K; s) \geq \wbar{h}^{(1)}(K; s) + \wbar{h}^{(2)}(K; s).\eqno{(4.1)}$$
\nl We shall now prove this theorem, which is in effect an optimized Weyl lower bound\sref{\weyl--\wein}; this remark will be clarified below, after the proof of the theorem. 
 From the definition (2.1) of kinetic potentials we have
$$\wbar{V}(K; s) = \inf_{{{\scriptstyle \psi \in {\cal D}(H)} \atop {\scriptstyle (\psi,\psi) = 1}} \atop {\scriptstyle (\psi, K\psi) = s}} (\psi, V\psi) 
= \inf_{{{\scriptstyle \psi \in {\cal D}(H)} \atop {\scriptstyle (\psi,\psi) = 1}} \atop {\scriptstyle (\psi, K\psi) = s}} \left(\psi, \left(h^{(1)} + h^{(2)}\right)\psi\right).$$
\nl But the latter minimum mean-value is clearly bounded below by the sum of the {\it separate}
 minima.  Thus we have  
$$\wbar{V}(K; s) \geq \inf_{{{\scriptstyle \psi \in {\cal D}(H)} \atop {\scriptstyle (\psi,\psi) = 1}} \atop {\scriptstyle (\psi, K\psi) = s}} \left(\psi, h^{(1)}\psi\right) 
+  \inf_{{{\scriptstyle \psi \in {\cal D}(H)} \atop {\scriptstyle (\psi,\psi) = 1}} \atop {\scriptstyle (\psi, K\psi) = s}} \left(\psi, h^{(2)}\psi\right) =
 \wbar{h}^{(1)}(K; s) + \wbar{h}^{(2)}(K; s),$$
\nl which inequality establishes the theorem.\qed

Another approach, which would eventually yield an alternative 
proof of the theorem, exhibits the relationship between Theorem~2 and the classical Weyl lower bound\sref{\weyl--\wein} for the eigenvalues of the sum of two operators.  Let us suppose that $\Psi$ is the exact normalized lowest eigenfunction of $H = K + V,$ so that $H\Psi = E\Psi.$  If the positive real parameter $w$ is bounded by $1,$ $0 < w < 1,$ then $E = (\Psi, (K + V)\Psi)$ may
be written as follows: 
$$\eqalign{E & =\  w\left(\Psi,\left(K + {1\over w}\ h^{(1)}(r)\right)\Psi\right)
+ (1-w)\left(\Psi, \left(K + {1\over{1-w}}\ h^{(2)}(r)\right)\Psi\right)\cr
& \geq\  w\inf_{{{\scriptstyle \psi \in {\cal D}(H)} \atop {\scriptstyle (\psi,\psi) = 1}}}
\left(\psi,\left(K + {1\over w}\ h^{(1)}(r)\right)\psi\right)\cr
&+\  (1-w)\inf_{{{\scriptstyle \psi \in {\cal D}(H)} \atop {\scriptstyle (\psi,\psi) = 1}}}
\left(\psi,\left(K + {1\over{1-w}}\ h^{(2)}(r)\right)\psi\right).}$$
\nl That is to say, in terms of component kinetic potentials, we arrive at Weyl's inequality
for the lowest eigenvalue of the sum $H = wK + h^{(1)} + (1-w)K + h^{(2)}:$ 
$$E \geq w\min_{s >0}\left\{s + {1\over w}\ \wbar{h}^{(1)}(K; s)\right\} + 
(1-w)\min_{s >0}\left\{s + {1\over{1-w}}\ \wbar{h}^{(2)}(K; s)\right\}.$$
\nl Since $w$ is an essentially free parameter in the last expression, we may optimize
the Weyl lower bound with respect
to the choice of $w$ and this forces the individual values of $s$ at the minima,
 $\{s_1(w), s_2(w)\},$ to be related.  More specifically we find from the
individual minimizations over $s,$
$$E \geq {\cal E}(w) = ws_1(w) + (1-w)s_2(w) + \wbar{h}^{(1)}(K; s_1(w))+ \wbar{h}^{(2)}(K; s_2(w)),$$
\nl where
$$w = -{{\partial\wbar{h}}\over{\partial s}}^{(1)}(K; s_1(w)),\quad{\rm and}\quad 1-w = -{{\partial\wbar{h}}\over{\partial s}}^{(2)}(K; s_2(w)).$$
\nl The critical condition ${\cal E}^{\prime}(w) = 0$ for the subsequent maximization of the lower bound over $w$ then yields $s_1(w) = s_2(w).$ 
Thus the best lower energy bound is 
given by
$$E \geq \min_{s >0}\left\{s + \wbar{h}^{(1)}(K; s)+ \wbar{h}^{(2)}(K; s)\right\}.$$
\nl The kinetic-potential inequality of Theorem~2 leads, of course, to the same energy lower bound: the optimization just performed above is therefore seen to be automatically `built in'
by the formalism. 

It follows immediately from the above kinetic-potential comparison theorem 
and coupling-parameter absorption that a lower bound to the lowest energy $E$ of
the Hamiltonian $H = K + \sum_{i}c_{i}h^{(i)}(r),$ $\{c_i >0\},$ is provided by the formula
$$E\  \geq\  \min_{s >0}\left\{s + \sum_{i}c_{i}\wbar{h}^{(i)}(K; s)\right\}.\eqno{(4.2)}$$
\nl Similarly we can extend this result to `continuous sums' such as 
$V(r) = \int_{t_1}^{t_2} c(t)h^{(t)}(r)dt.$

  This general theory becomes practically 
useful when we have good information concerning the components.  More particularly, we must have some exact component kinetic potentials, or lower bounds to them.  Outside the well-explored Schr\"odinger
case $K = p^2,$ such analytical results are rather sparse. We look at the interesting
class of power-law potentials in the next section.
\title{5.~~Sums of powers and the log potential} 
For power-law
potentials and the relativistic kinetic energy $K = \sqrt{m^2+p^2}$ we have discussed 
some lower bounds in Section~3 and we shall now turn these to our advantage. 
The link between Theorem~1 and Theorem~2 derives from the observation
that the equation $K(s) = \sqrt{m^2 + s^2}$ allows us to change the minimization
variable $s\rightarrow r = 1/s.$ In the first stage of minimization, we have used
Jensen's inequality for the lower bound (see proof of Theorem~1); this allows us to keep $(\psi, p\psi) = s = 1/r$ constant at first, and then later minimize over $s,$ or, equivalently, over $r.$  We can also easily accommodate a further positive kinetic-energy
parameter $\beta.$ Thus we immediately arrive at
\nll{\bf Theorem~3}~~
{\th A lower bound to the lowest eigenvalue of the
semirelativistic spinless-Salpeter operator
$$H = \beta\sqrt{m^2 + p^2} + \sum_{q\neq 0}a(q)\sgn(q)r^q + a(0)\ln(r),$$
\nl where $\beta > 0$ and the potential coefficients $a(q) \geq 0$ are not all zero,
 is given by
$$E\ \geq\  \min_{r > 0}\left\{\beta\sqrt{m^2 + {1\over{r^2}}} + 
\sum_{q\neq 0}a(q)\sgn(q)(P^{(1)}(q)r)^q + a(0)\ln(P^{(1)}(0)r)\right\},\eqno{(5.1)}$$
where, for the Coulomb component $q = -1,$ we 
make the substitution}
$$-{a(-1)\over{P^{(1)}(-1)r}} = -{{\beta^2 mv}\over{e_L(v)r}} = -{{\beta v}\over r}\left({2\over {1 + \sqrt{1-4v^2}}}\right)^{\half},\quad v = {{a(-1)}\over\beta} < \half.\eqno{(5.2)}$$  
\nl The problem presented for the lower bound by the fact that $p -v/r$ has no
discrete spectrum was discussed in Section~2. We have no simple \hi{P}{number} $P(-1)$ but we could derive
a `running' $P$ (2.13) from the Martin--Roy energy bound (2.11); the positive factor $\beta$ has
been inserted in (5.2) by elementary scaling.  We shall look at applications
of  Theorem~3 in Section~7 when we also have at our disposal the upper-bound \hi{P}{numbers}
derived in Section~6.
\title{6.~~Variational upper bounds} 
The lower bound for sums discussed in the previous two sections has the 
attractive feature that if the component kinetic potentials are exact
and only one term is present, then the result is exact.  We are unable to
construct a general upper bound with this feature.  Instead we use a trial wave
function $\phi = c\exp(-\half\alpha r^\nu)$ with a scale parameter $\alpha>0$ 
and two other parameters $\{c, \nu\},$ and we
apply this wave function to the entire problem.  One degree of freedom $c$ is
used to guarantee normalization, and the scale parameter $\alpha >0$ is expressed
in terms of a new variable $t > 0$ chosen in such a way that the scale minimization is
of an expression with the same form as the lower bound. Initially we use here $t$ rather than 
$r$ since, during the discussion, we shall need to refer to the
 potential function $V(r).$ The choice of
the remaining parameter $\nu > 0$ is left for later optimization.

If we suppose that $c$ has already been chosen so that $\|\phi\| = 1,$ and,
 for computational convenience, we use Jensen's inequality, we then obtain the following
upper energy bound: 
$$E < {\cal E} = \beta\sqrt{m^2 + (\phi, p^2 \phi)} + 
\left(\phi,\left\{\sum_{q\neq 0}a(q)\sgn(q)r^q + a(0)\ln(r)\right\}\phi\right).\eqno{(6.1)}$$
Now, for each fixed $\nu>0,$ we define a new scale variable $t>0$ by the following:
$$(\phi, p^2\phi) = (\phi,-\Delta\phi)  = \alpha^{2/\nu}\left({\nu\over 2}\right)^2 
{{\Gamma(2+{1\over \nu})}\over{\Gamma({3\over \nu})}} \equiv {1\over t^2}.\eqno{(6.2)}$$
\nl Using this definition of $t$, we can go on to define the `upper' \hi{P}{numbers} 
${\cal P}(\nu,q)$ by the relations
$$(\phi, r^q\phi) = {{1}\over {\alpha^{q/\nu}}}{{\Gamma({{q+3}\over \nu})}\over{\Gamma({3\over \nu})}}
\equiv \left({\cal P}(\nu,q)t\right)^q,\quad q \ne 0,\eqno{(6.3a)}$$
and
$$(\phi, \ln(r)\phi) = \ln\left({\cal P}(\nu, 0)t\right).\eqno{(6.3b)}$$
\nl If we now rename the scale variable $t = r,$ and minimize the upper bound 
${\cal E}$ with respect to scale, we arrive at   
\nll{\bf Theorem~4}~~
{\th For each $\nu > 0,$ an upper bound to the lowest eigenvalue $E$ of the
Salpeter operator
$$H = \beta\sqrt{m^2 + p^2} + \sum_{q\neq 0}a(q)\sgn(q)r^q + a(0)\ln(r),$$
\nl where $\beta > 0$ and the potential coefficients $a(q) \geq 0$ are not all zero, is given by
$$E\  \leq\  \min_{r > 0}\left\{\beta\sqrt{m^2 + {1\over{r^2}}} + 
\sum_{q\neq 0}a(q)\sgn(q)({\cal P}(\nu,q)r)^q + a(0)\ln({\cal P}(\nu,0)r)\right\},\eqno{(6.4)}$$
where the upper \hi{P}{numbers} are provided by the formulas
$${\cal P}(\nu,q) = {\nu\over 2}\left({{\Gamma(2+{1\over \nu})}\over{\Gamma({3\over \nu})}}\right)^{\half}
\left({{\Gamma({{q+3}\over \nu})}\over{\Gamma({3\over \nu})}}\right)^{{1\over q}},\quad q\neq 0,\eqno{(6.5a)}$$
$${\cal P}(\nu,0) = {\nu\over 2}\left({{\Gamma(2+{1\over \nu})}\over{\Gamma({3\over \nu})}}\right)^{\half}
\exp\left({1\over \nu}\psi\left({3\over \nu}\right)\right),\eqno{(6.5b)}$$
\nl and $\psi$ is the digamma function $\psi(t) = \Gamma'(t)/\Gamma(t).$
}
\nll Apart from the special Coulomb considerations pertaining to the lower bound (5.1),
 that formula is essentially identical to the upper bound (6.4):
 we simply have to use the correct \hi{P}{numbers} in each case.   
\title{7.~~Examples} 
We have now assembled the \hi{P}{numbers} for our energy-bound formulas (5.1) and (6.4).
We shall use the lower \hi{P}{numbers} in Table~1,
lower `running' \hi{P}{formula} for the Coulomb component (2.13),
and the formulas (6.5) for the ${\cal P}(\nu,q)$ corresponding to the 
variational upper bound (6.4).  The class of problems we are thus immediately able to
consider have the following explicit Hamiltonian form:
$$H = \beta\sqrt{m^2 + p^2} - a/r + b\ln(r) + c r + d r^2,\quad
 a,b,c,d \geq 0,\eqno{(7.1)}$$
\nl where $\beta > 0,$ and the potential parameters $\{a,b,c,d\}$ are not all zero.
We look at two examples.  In the first, illustrated in Figure~2, we look at the linear-plus-Coulomb potential $V(r) = -0.1/r + 0.25 r$ 
and compare the energy bounds $\{L,~U\}$ we find, as functions of the mass $m,$ with some very accurate numerical values (center curve) obtained by minimizing the expectation value of the Hamiltonian in a \hi{25}{dimensional}
trial space.  In the next graph, Figure~3, we plot the energy bounds alone, for the 
same potential and a wider range of values of the mass $m.$  As a second example we consider the
broad linear combination $V(r) = -0.1/r + 0.25 \ln(r) +0.25 r + 0.25 r^2$ and plot in Figure~4 the
energy bounds as functions of the mass.  These illustrations give a clear indication of the quality of the bounds that the theory yields. 
\title{8.~~Conclusion} 
The principal theoretical results of this paper are the complementary bounds of Theorem~1,
and the sum-approximation lower bound, Theorem~2. In order to arrive at these results we needed
first to extend the notion of `kinetic potential' to allow for more general kinetic-energy operators than the Schr\"odinger form $K = p^2.$  The complementary bounds are based on the assumption that $K$ is a convex function of $p$ and also a concave function of $p^2,$ assumptions 
clearly satisfied by our prime example and principal motivation, 
the relativistic kinetic energy $K = \beta\sqrt{m^2 + p^2}.$  The inequality of Jensen then allows us
to learn approximately how special mean values of the problem, the eigenvalues of $H,$ depend on the operator parameters.

By combining Theorem~1 of this paper with Theorem~2 of our earlier paper\sref{\hallsa} we obtain a general theory applicable to `operator manifolds' of the form 
 $H = K(p) + g(h)$ with, on the one hand, 
tangent spaces spanned by the Schr\"odinger operators $ap^2 + b h(r) + c,$ 
and, on the other, by complementary operators of the form $ap + b h(r) + c.$ Given the
correct convexities of $K$ and $g,$ energy bounds immediately follow.
We looked at one example of this type of problem near the end of Section~3; and the results were exhibited in Figure~1.

 A completely different lower bound is provided by Theorem~2,
which may be thought of as a spectral expression of the 
sum structure of the potential, namely the subadditivity of the
corresponding kinetic potential, as a sum of components.

In order to make practical use of these theoretical results we need some definite
spectral information about component problems.  This is provided by the family of 
pure-power potentials $V(r) = \sgn(q) r^q.$  For this family we are able to take 
advantage of known eigenvalues, or bounds to them, and of simple upper bounds obtained
with the aid of Jensen's inequality and a two-parameter family of trial functions.
All our component results can then be expressed in terms of certain \hi{P}{numbers} 
(or, for the lower Coulomb case, $q= -1$, by a \hi{P}{function}), which are required by the
general lower- and upper-bound formulas of Theorems~3 and 4. 
 These formulas illustrate the effectiveness of the theoretical results and provide
recipes for approximate solutions to an interesting class of semirelativistic spectral problems.   
\title{Acknowledgements} 
  Partial financial support of this work under Grant No. GP3438 from the Natural 
Sciences and Engineering Research Council of Canada, and hospitality of the 
Institute for High Energy Physics, and
of the Institute for Theoretical Physics, of the Austrian Academy of Sciences, in Vienna,
 is gratefully acknowledged by one of us [RLH].\medskip 
\references{1} 
\np
\nl {\bf Table 1.}~~Eigenvalues for $v = 1$ and corresponding \hi{P}{numbers}
 [given by (2.6) and (2.10)] for the Coulomb, log, linear and harmonic-oscillator potentials. The eigenvalues have been computed numerically and are rounded so that the $E^{(1)}(q)$ are lower bounds and the $E^{(2)}(q)$ are upper bounds (and similarly for the derived \hi{P}{numbers}).
The Coulomb lower bound is treated differently because $H=p-1/r$ has
no discrete spectrum.\medskip  


$$\vbox{\offinterlineskip
 \def\vr{\vrule height 24 true pt depth 12 true pt}
 \def\vra{\vr\hfill} \def\vrb{\hfill &\vra} \def\vrc{\hfill & \vr\cr\hrule}
 \def\vrq{\vr\quad} 
\hrule
\settabs
\+ \vrq \kern 0.5true in\ &\vrq \kern 1true in &\vrq \kern 1true in 
&\vrq \kern 1true in &\vrq \kern 1true in &\cr\hrule

\+ \vra $q$ \vrb $E^{(1)}(q)$ \vrb $P^{(1)}(q)$\vrb $E^{(2)}(q)$\vrb $P^{(2)}(q)$\vrc
\+ \vra $-1$ \vrb --- \vrb --- \vrb $-{1\over 4}$ \vrb $1$\vrc
\+ \vra $0$ \vrb $1.06365$ \vrb $1.0657$ \vrb $1.0443325$ \vrb $1.218669$\vrc
\+ \vra 1 \vrb $2.23225$ \vrb $1.2457$ \vrb $2.3381075$ \vrb $1.376084$\vrc
\+ \vra $2$ \vrb $2.338107$ \vrb 1.366687 \vrb $3$ \vrb ${3\over 2}$\vrc

}$$

\baselineskip 18 true pt 
\np
\hbox{\vbox{\psfig{figure=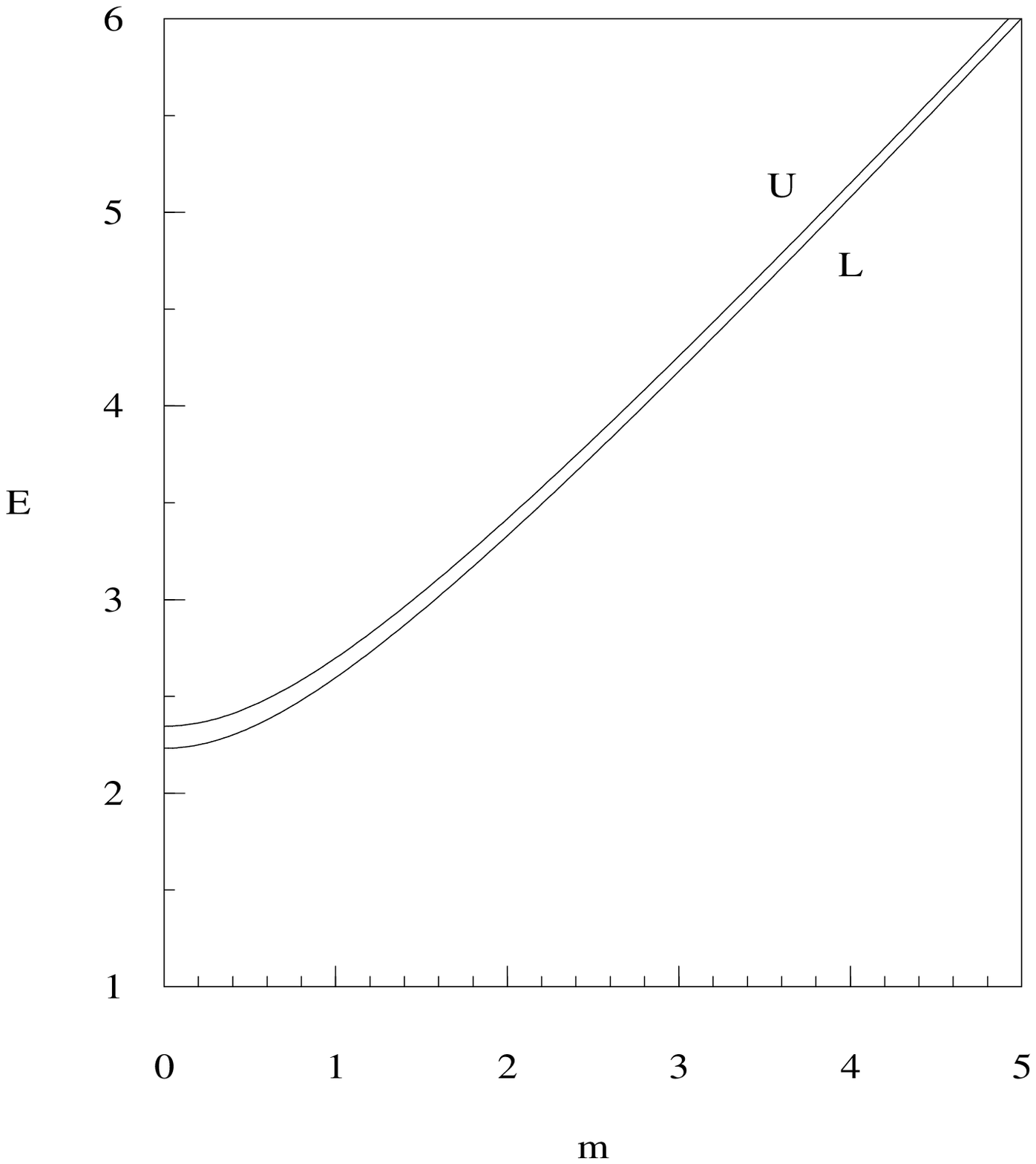,height=6in,width=5in,silent=}}}

\title{Figure 1.}
\nl Complementary upper ($U$) and lower ($L$) bounds (3.4) on the lowest eigenvalue $E(m)$
 of $H = \sqrt{m^2 + p^2} + r$ plotted against $m$.
\np
\hbox{\vbox{\psfig{figure=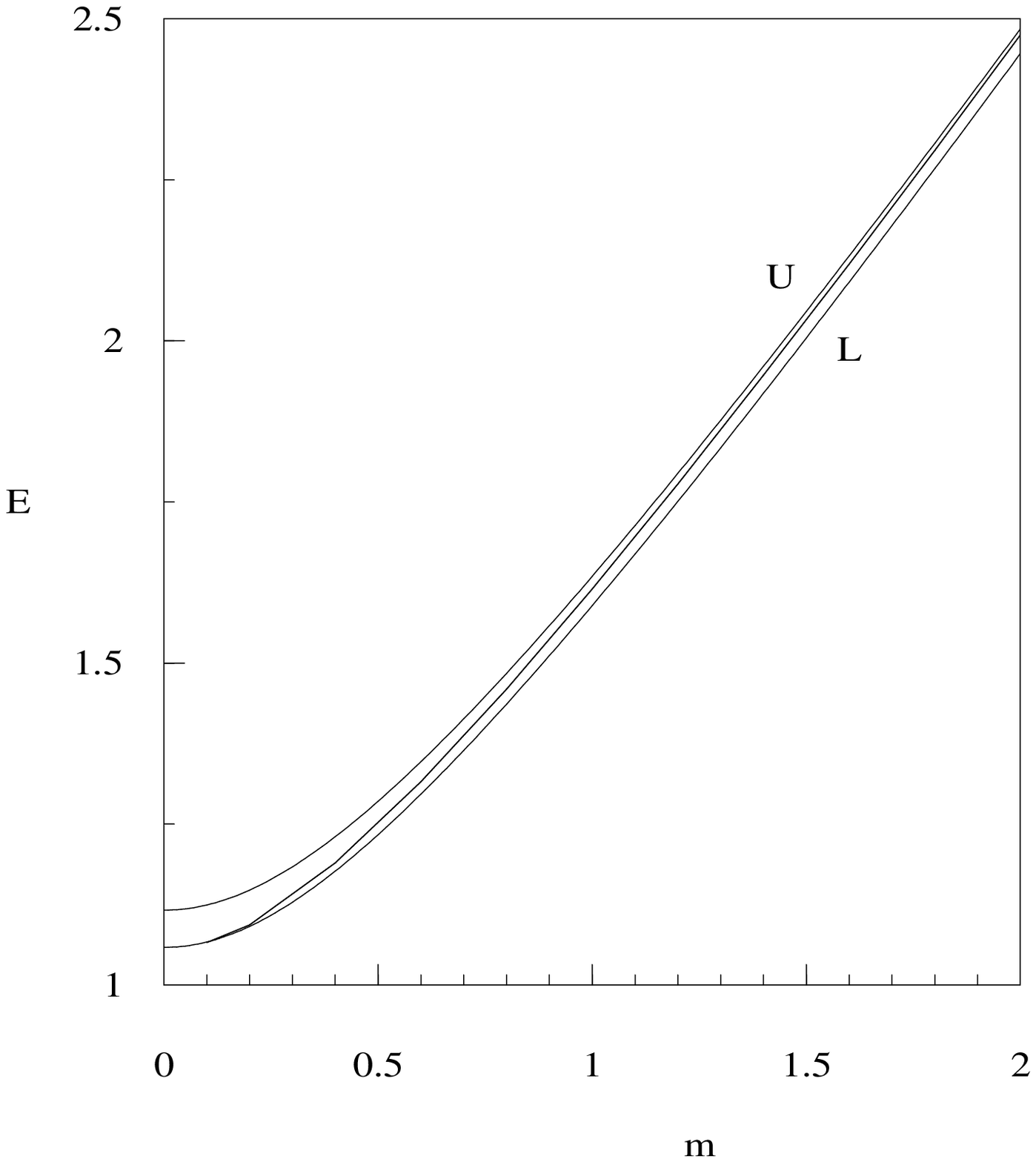,height=6in,width=5in,silent=}}}
\title{Figure 2.}
\nl Lower bounds ($L$) by (5.1) and upper bounds ($U$) by (6.4) for the lowest 
eigenvalue $E(m)$  of $H = \sqrt{m^2 + p^2} -0.1/r + 0.25 r$ 
plotted against $m$.  The upper bound ($U$) used the wave-function parameter $\nu = 1.6.$ 
The central curve is a very accurate upper bound found by a variational exploration in a \hi{25}{dimensional} trial space.  
\np
\hbox{\vbox{\psfig{figure=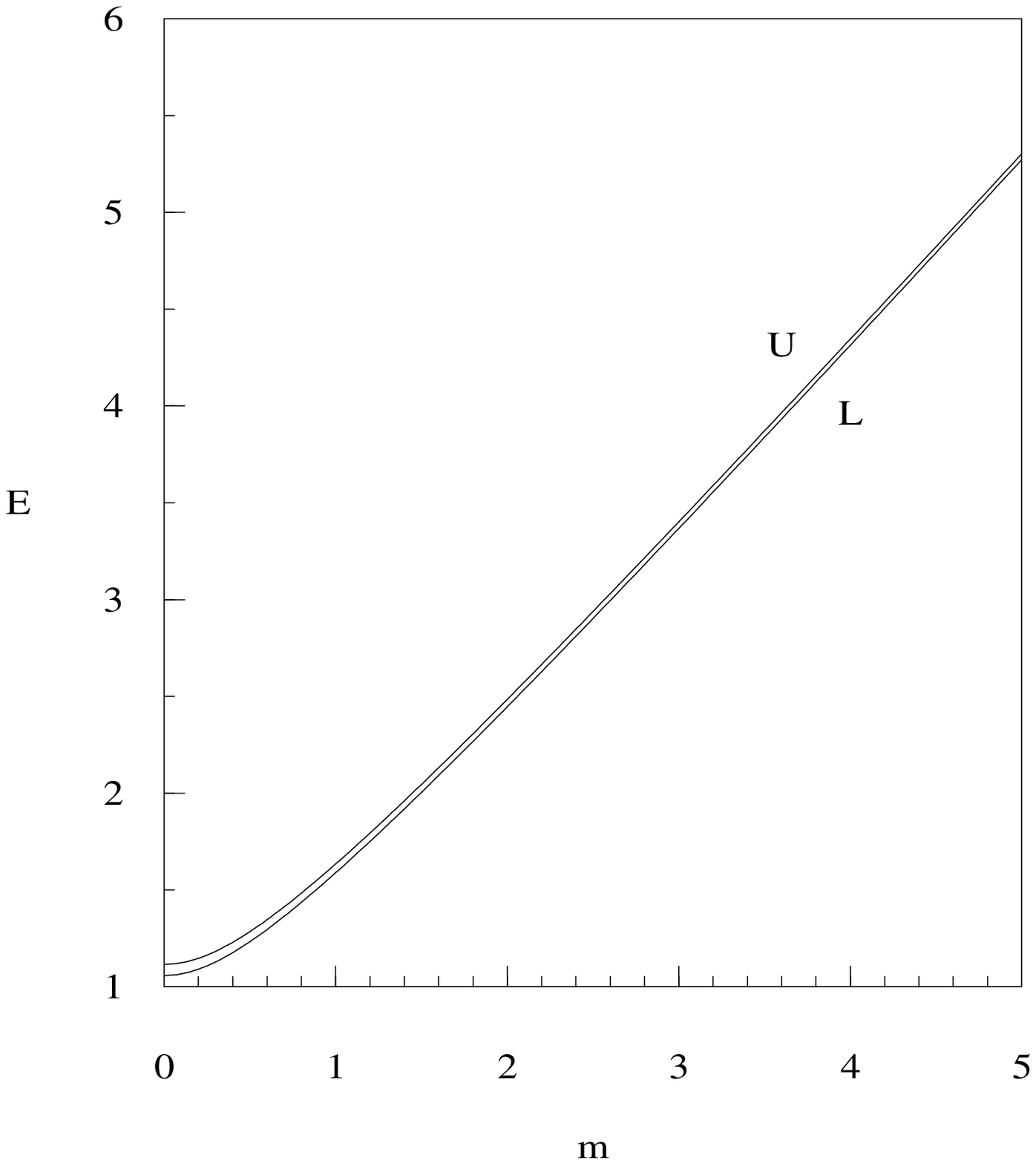,height=6in,width=5in,silent=}}}
\title{Figure 3.}
\nl Lower bounds ($L$) by (5.1) and upper bounds ($U$) by (6.4) for the lowest 
eigenvalue $E(m)$  of $H = \sqrt{m^2 + p^2} -0.1/r + 0.25 r$ 
plotted against $m$: this is a continuation of the graph in Figure~2 to larger $m.$ 
\np
\hbox{\vbox{\psfig{figure=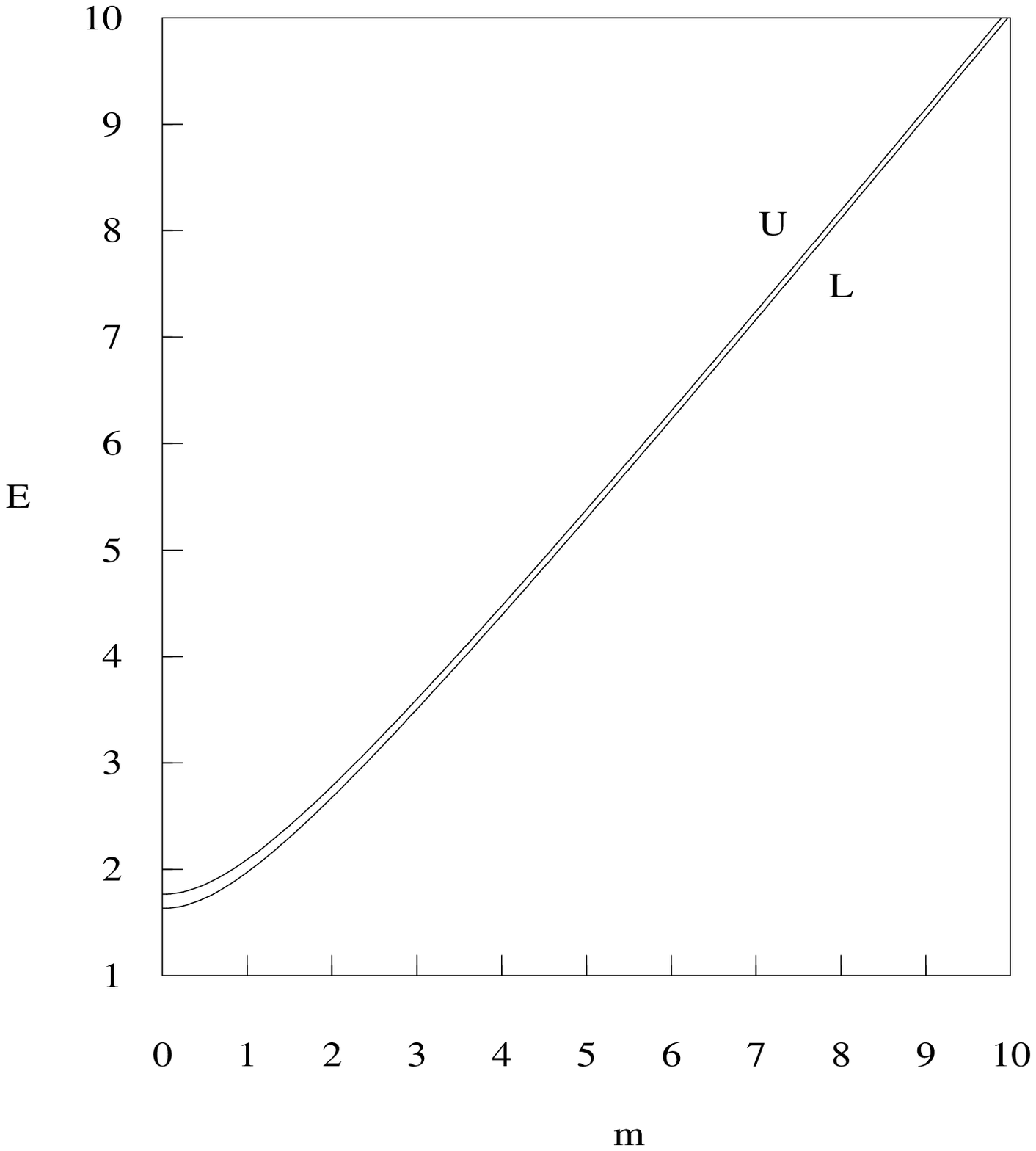,height=6in,width=5in,silent=}}}
\title{Figure 4.}
\nl Lower bounds ($L$) by (5.1) and upper bounds ($U$) by (6.4) for the lowest 
eigenvalue $E(m)$ of $H = \sqrt{m^2 + {\bf p}^2} -0.1/r + 0.25 \ln(r) +0.25 r + 0.25 r^2$ 
plotted against $m$. The upper bound ($U$) used the wave-function parameter $\nu = 1.4.$

\end